\documentclass[12pt]{article}
\usepackage{graphicx}
\usepackage{color}
\newcommand{\cit} [1] {\mbox{$^{\cite{#1}}$}}

\newcommand{\sub} [2] {\mbox{$#1_{\mbox{\scriptsize{#2}}}$}}

\newcommand{\qot}{\mbox{$\sub{Q}{0}(t)$}}

\newcommand{\degrees}{\mbox{$^\circ $}}

\begin{document}
\bibliographystyle{unsrt}	

%
%
\begin{titlepage}
\vspace{.2in}
\centerline{\large\bf Charge Offset Stability in Si Single Electron Devices with Al Gates}
\vspace{.1in}
\centerline{Neil M. Zimmerman}
\vspace{.2in}
\centerline{+1-301-975-5887}
\centerline{neilz@mailaps.org; ftp://ftp.nist.gov/pub/physics/neilz/papers.html}
\centerline{National Institute of Standards and
Technology, Gaithersburg, MD 20899, USA}
\vspace{.2in}
\centerline{Chih-Hwan Yang, Nai Shyan Lai$^*$, Wee Han Lim$^*$, and Andrew S. Dzurak}
\vspace{.2in}
\centerline{ARC Centre of Excellence for Quantum Computation and Communication Technology,}
\centerline{School of Electrical Engineering and Telecommunications}
\centerline{The University of New South Wales, Sydney 2052, Australia}
\centerline{$^*$ now at Asia Pacific University of Technology and Innovation,}
\centerline{Technology Park Malaysia, 57000 Bukit Jalil, Malaysia}
\vspace{.2in}
\centerline{\today}

\renewcommand{\baselinestretch}{1.5}

\begin{abstract}

We report on the charge offset drift (time stability) in Si single electron devices (SEDs)
defined with aluminum (Al) gates.  The size of the charge offset drift (0.15 $e$) is intermediate
between that of Al/AlO$_x$/Al tunnel junctions (greater than 1 $e$) and Si SEDs
defined with Si gates (0.01 $e$).  This range of values suggests that defects in the AlO$_x$
are the main cause of the charge offset drift instability.

\end{abstract}
\end{titlepage}

%
%

%
\large\renewcommand{\baselinestretch}{1.5}\normalsize

Single-electron devices (SEDs) have been proposed for a variety of applications, including
electrical metrology (standards of current and charge)\cit{Zimmerman03b}, ultra-sensitive
sensors including charge electrometers\cit{Likharev99a}, and as solid-state
qubits\cit{Petta05a}.  Si-based single electron devices are one of the leading candidates
for these applications, in part because of their general attractive attributes including
tunability\cit{Fujiwara06a}, compatibility with present-day integrated circuits, and
because of their stability (lack of charge offset drift)\cit{Zimmerman08a}.  Referring
specifically to their potential as spin qubits: The weak spin-orbit coupling and low
density of nuclear spins in naturally occurring silicon means that it is ideally suited as
a host for spin qubits\cit{Morton11a}, with recent demonstrations of electron spin
qubits\cit{Maune12a}\cit{Pla12a} and a high-fidelity nuclear spin
qubit\cit{Pla13a}. Furthermore, the coherence time in bulk Si can be made very long when
the nuclear spin bath is effectively removed through the use of isotopically-enriched
$^{28}$Si\cit{Tyryshkin12a}\cit{Steger12a}.

One of the important attributes for all of the applications mentioned above is the time
stability of the SEDs.  This is a particular issue because the inherent sensitivity to the
motion of a single electron has both attractive and deleterious implications: it is
attractive because SEDs provide the world's most sensitive charge electrometers; it is
deleterious because their gross behavior can be markedly changed by small subtle movements
of nearby charges.  These devices are fabricated with thin-film lithography and processing
on the surfaces of substrates; thus, as opposed to bulk single-crystal Si, in these
devices there are numerous nearby defects which can possess a net charge or dipole moment.  In
turn, these charges can modulate the electrostatic potential of the SED island, and thus
lead to a random time instability.  This manifests itself as a time-dependent random phase
offset $\phi$\cit{Zimmerman08a} to the periodic control curve (e.g., the inset to Figure
2), quantified as \qot $= \phi/2\pi$ $e$, where $e$ is the electron charge.

In addition to the effect on the prospects for integration, the potential application of SEDs as
qubits gives additional impetus to the importance of assessing the time stability.  It is
generally believed that electron quantum coherence is more "fragile", i.e., more prone to
loss of information, than classical storage, in part due to the effect of nearby defects that can have
random fluctuations of their charge or spin.  Thus, similar to studies of properties such
as electron mobility\cit{Nordberg09a}, elucidation of the charge offset drift \qot\ can give us
additional information as to the suitability of particular materials or device
architectures with regard to optimizing fidelity and coherence in qubit devices.

In previous work (\cit{Zimmerman08a} and references therein), we have shown that there is
a marked difference between \qot\ in metal devices (based on Al/AlO$_x$/Al tunnel
junctions) and Si-based devices containing only crystalline and polycrystalline Si and
SiO$_2$: the typical amplitude of \qot\ is about 0.01 $e$ in the Si-based devices and
greater than 1 $e$ in the metal devices.  Recently, Si-based devices with Al gates have been
shown to have excellent behavior in a variety of applications\cit{Lim09a} including spin
qubit coherence\cit{Pla12a}.  The natural question thus arises, in the context of previous
work on \qot: Does the presence of Al gates affect the charge offset drift in Si devices?
We aim to answer that question in this paper.

Our fabrication (Figure 1a and sketch in Table left column) followed closely previous
work\cit{Lai11a}\cit{Huebl10a}.  Starting from a high-resistivity (10 k$\Omega$-cm n-type)
Si substrate, we generated by diffusion highly doped n$^+$ source (S) and drain (D) ohmic
contacts, and then grew a thermal gate oxide (18 nm of SiO$_2$) at 800\degrees\ C in O$_2$ and
dichloroethylene.  We then fabricated a three-layer gate stack: i) Al barrier gates B1 and
B2 plus a AlO$_x$ isolation oxide, ii) Al lead gates L1 and L2 and another AlO$_x$
isolation oxide, and then finally iii) the Al plunger gate P.  Lead gates L1 and L2
terminate slightly inside B1 and B2; P fills the length between B1 and B2.  Thus, in
various locations (see Fig 1a), the stack can have one (e.g., far away from center), two
(e.g., where P lies over L1 to left of B1) or three (e.g., on top of B1) Al layers.

The gates were all formed by electron beam lithography and lift-off patterning of
thermally-evaporated aluminum.  The isolation oxides were formed in air at 150\degrees\ C,
resulting in about 4 nm of AlO$_x$.  Finally, we annealed in forming gas (15 minutes,
400\degrees\ C, 5\% H$_2$), followed by cleaving and wire bonding for electrical contact.
Figure 1b is a cross-sectional TEM micrograph of the finished device directly underneath
gate P (Al metal), showing among other details the undeliberate formation of a thin
interfacial layer of AlO$_x$ between the SiO$_2$ and the Al gate.

As depicted in the schematic circuit in Figure 1a, we applied a small drain voltage to ohmic
contact D, measured the current flowing through ohmic contact S, used \sub{V}{L1} and
\sub{V}{L2} to induce a conducting accumulation layer between S and D at the Si/SiO$_2$
interface, and generated tunneling barriers by applying mildly negative (with respect to
the threshold voltage) voltages \sub{V}{B1} and \sub{V}{B2}.  The combination of these
produced a quantum dot at the center of the device, whose chemical potential we controlled
with \sub{V}{P}.  We applied voltages and measured the current using commercial voltage
sources and current amplifiers.  All the measurements presented in this paper were
performed at 2.2 K in vacuum in a cryocooled measurement system.

The inset to Figure 2 shows the standard Coulomb blockade oscillation (CBO); the peak spacing is
constant over a fairly large number of oscillations, and shows an overall mild monotonic
increase in current as increasing \sub{V}{P} lowers the height of the tunnel barriers, and
thereby increases the current.  In order to measure the time stability \qot, we repeatedly
make measurements of the CBO, and for each measurement fit\cit{Zimmerman08a} $ \sub{I}{D}(
\sub{V}{P}) = A_0 + A \sin[2\pi(\sub{V}{P}/\Delta \sub{V}{P} + \qot)] + B \sub{V}{P}$.
Here, $A_0$ is a current offset, $A \approx 0.1$ nA, $\Delta \sub{V}{P} \approx 22$ mV
is the period, and $B \approx 0.4$ pA/mV accounts for the mild linear slope as seen in Figure 2 inset.

The uncertainty in the measurement, arising from the uncertainty of the fit, is about
$\pm$ 0.01 $e$; the sample-dependent fluctuation in \qot\ occurs on time scales of about 0.1
days and greater, and yields a total range in \qot\ of about $\pm$ 0.15 $e$.  We measured
identical behavior for two different nominally identical devices using two different sets
of measurement electronics and ramp protocols, and also verified the accuracy of the
measurement by demonstrating a much smaller drift of about $\pm$ 0.01 $e$ in a
Si/poly-Si device\cit{Koppinen13a} with the same measurement system and temperature.

To put this in context, in earlier work we noted that the typical
amplitude\cit{Zimmerman08a} of \qot\ in metal SEDs (based on Al/AlO$_x$/Al tunnel
junctions) is greater than 1 $e$, and in Si-based devices containing only crystalline and
polycrystalline Si and SiO$_2$ the amplitude is about 0.01 $e$.  In this earlier work, we
demonstrated that the reason for this difference in the behavior of \qot\ was due to the
instability of the AlO$_x$ as opposed to the SiO$_2$.  In particular, the time-dependent
fluctuators which give rise to \qot\ exist in both insulators, but interactions between
the fluctuators in the AlO$_x$ also give rise to a glassy relaxation and thus to the time
instability in \qot.  In order to make sense of the present results, we focus on oxide
similarities and differences between the CMOS-compatible\cit{aside_CMOS} SEDs and the
devices studied in this paper (see Table 1).

From the Table we note the following correlations between amplitude of \qot\ and device characteristics:

\begin{description}

	\item[Presence of AlO$_x$ (AlO$_x$ thickness)] As discussed above.
	\item[Total thickness of oxide t] Based on two data points (Si/SiO$_2$/Al gates
          versus Al/Alsub{O}{x}/Al), it appears that smaller total oxide thickness is
          correlated with a larger amplitude of \qot.  This is consistent with a simple
          estimate for the change in charge displacement on the quantum dot as a function
          of oxide thickness (see below).
	\item[Electric field strength in AlO$_x$ (AlO$_x$ E)] The fact that the electric
          field strength is smaller for the largest amplitude of \qot\ indicates that the
          applied voltage is not inducing the drift, and might in fact be inhibiting it.
          This is consistent with a previous observation of instability as a function of
          gate voltage in our devices\cit{Yang13a}.
	\item[Current through AlO$_x$ (AlO$_x$ \sub{I}{D})] The devices with the largest
          amplitude of \qot\ are the only ones in which AlO$_x$ exists in the tunnel
          barriers, and therefore in which the AlO$_x$ current $\sub{I}{D} \neq 0$.  This
          suggests that, among other things, electromigration might be a contributing
          factor to the charge offset drift (see below).

\end{description}

In addition to the previous discussion of the instability of AlO$_x$, we also note a
previous suggestion\cit{Brown06a} that large charge offset fluctuations in a Al/AlO$_x$/Al
SED were due to isolated Al grains.  These grains were generated during deposition from
thermal evaporation, and were identified by scanning electron micrographs.  The hypothesis
is that such grains act as sources/sinks of charge which randomly fill and empty on a
variety of time scales, depending on the tunneling resistance between the isolated grains.

Electromigration\cit{electromigration_book} is a well-known driving source of atomic
motion in microscale (e.g., integrated circuit metallization) and nanoscale (e.g., single
atom junctions) conductors.  It is generally believed to arise from both electric
field-induced motion of the massive ions and from momentum transfer from hot electrons.
It has been previously observed\cit{Zimmerman97b} that a particularly high level of charge
offset drift in a Al/AlO$_x$/Al SED appeared to be driven by current through the junction.
On the other hand, a compendium\cit{Zimmerman08a} of \qot\ results in Al/AlO$_x$/Al SETs
showed no correlation between \sub{I}{D} and amplitude of \qot.  Thus, it appears
plausible that in some cases the charge offset drift in Al/AlO$_x$/Al
is due to electromigration, but certainly not in all cases.

A simple estimate\cit{Song95a} for the change in charge displacement $\Delta Q_0$ can be
derived as follows: For a bare charge of magnitude {\it e} moving a perpendicular distance
$d$ in a parallel plate capacitor with insulator thickness $t$, $\Delta Q_0 \approx d/t$
$e$; for typical values ($d$ an interatomic distance, $t$ a few nm), this leads to $\Delta
Q_0 \approx 0.1$ $e$.  If we consider a charge dipole with change in perpendicular dipole
length $\Delta l$, we obtain $\Delta Q_0 \approx \Delta l/t$ $e$, which will yield a
somewhat smaller but similar magnitude.  These estimates (valid for both gate and barrier
insulators) also indicate that the charge offset drift amplitude should scale inversely
with insulator thickness.

In contrasting the influence of electric field versus oxide thickness, we can point out
that in previous work\cit{Zorin96a} in which the authors measured and modelled the noise
as a function of local position, the devices studied had AlO$_x$ only in the tunnel junctions, and
not between the gate and island.  For this reason, considering various insulators
surrounding the dot, the electric field strength $E \propto 1/t$ and thus was correlated
with insulator thickness.  In our case, the addition of a vertically-located gate allowed us to
discriminate between the effects of applied voltage and insulator thickness.

We thus reach the following conclusions from our present work:

\begin{enumerate}

	\item Al gates on top of SiO$_2$ result in a thin interfacial layer of
          AlO$_x$\cit{Lim09a} (see Figure 1b),
          important because it is the only AlO$_x$ which is not electrostatically screened
          by Al gates.
	\item The amplitude of \qot\ in devices containing AlO$_x$ and SiO$_2$ is
          larger than devices containing only SiO$_2$.  This is because of either: (i)
          inherent glassiness of atomic/molecular motion in the AlO$_x$, or (ii) separated
          Al grains at the edges of the gates.
	\item The fact that \qot\ is smaller for Si/SiO$_2$/Al devices ($t = 20$ nm) than
          for Al/Al\sub{O}{x}/Al) devices ($t = 2$ nm) is consistent with a very simple
          model that predicts the dependence of \qot\ amplitude upon
          distance; this suggests that moving Al\sub{O}{x} layers
          further away from device layers may be very helpful in
          reducing \qot.

\end{enumerate}

Finally, we can comment on the consequences of our work for future devices.  One important
goal of single electron metrology\cit{Zimmerman03b} is that of a single electron current
standard with large value, where one approach is to parallelize a large number of devices.
In such parallelized devices, the absence of charge offset drift would be evidently
important.  For quantum information, a significant candidate for solid-state qubits is
spins in Si\cit{Morton11a}.  Silicon SEDs with the same materials and device architecture
as those studied here have recently been used to read out the state of an electron spin
qubit bound to a nearby phosphorus ($^{31}$P) donor\cit{Morello10a}.  Such SED-donor
coupled systems have been used in demonstrations of coherent control of both electron
spin\cit{Pla12a} and nuclear spin\cit{Pla13a} qubits. In the case of the electron spin and
the ionised $^{31}$P nuclear spin the coherence times appear to be determined by the
dynamics of the $^{29}$Si nuclear spin bath present in natural Si. For the neutral
$^{31}$P nuclear spin, however, the coherence time appears to be limited by an additional
mechanism.  In combination with the present work, since \qot\ is the manifestation of
chemical potential fluctuations and since defect fluctuations occur on a broad
distribution of timescales, charge noise from the AlO$_x$ layers surrounding the Al gates
could well be one possibility.  It thus appears that by ameliorating or eliminating the
effect of Al and AlO$_x$ in our devices, we may be able to improve the coherence times in
future experiments.  Finally, looking further in the future towards large-scale quantum
processors, the use of CMOS-compatible\cit{aside_CMOS} device architectures would appear
to be a sensible approach to avoid the deleterious effects of charge noise.

We are grateful to acknowledge useful conversations with Justin Perron, Alessandro Rossi,
Andrea Morello, Michael Stewart, Jr., and Ted Thorbeck, as well as the microscopic
technical assistance of Joshua Shumacher and Alline Myers.  Research was performed in part
at the NIST Center for Nanoscale Science and Technology (CNST).  C.H.Y. and
A.S.D. acknowledge support from the Australian National Fabrication Facility, the
Australian Research Council (under contract CE110001027), and the U.S. Army Research
Office (under contract W911NF-13-1-0024).



%
%
\vfil\break
\centerline{List of Figures and Tables}

\begin{figure}[htbp]
	\centering
	\includegraphics[width=1\textwidth]{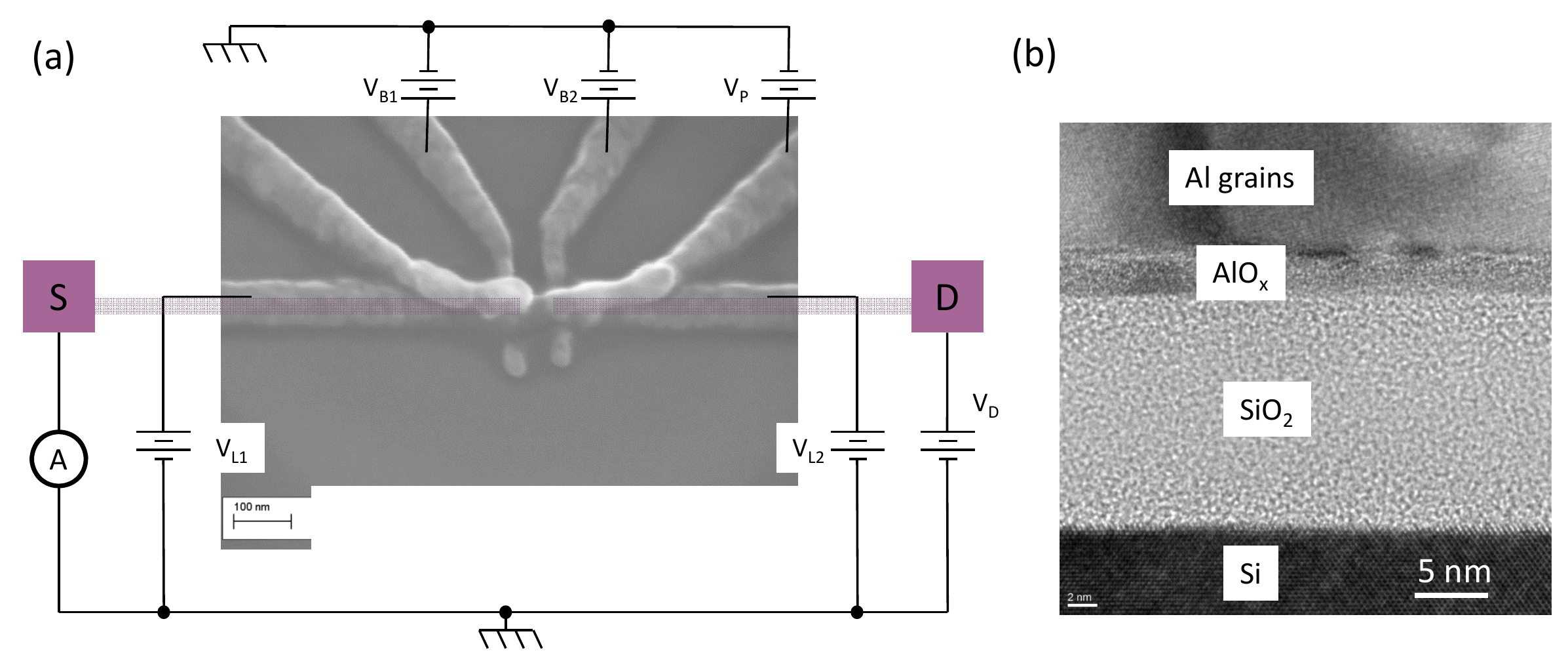}
	\caption[Figure 1]{
        (a) Scanning electron micrograph of Al gate electrodes on Si/SiO$_2$ substrate
        (30\degrees\ tilt) and schematic measurement circuit.  The solid pink (grey on paper)
        squares labelled ``S'' and ``D'' indicate the heavily-doped n$^+$ source and drain
        regions, and the transparent pink rectangle schematically indicates the conducting
        accumulation layer (generated by \sub{V}{L1} and \sub{V}{L2}) in the Si at the Si/SiO$_2$
        interface.
        
        (b) Cross-sectional transmission electron micrograph taken in the middle of the device in
        Figure 1a , showing both i) the general thin-film stack and
        specifically ii) the thin AlO$_x$ layer underneath the aluminum gate.}
	\label{fig:SEM_TEM}
\end{figure}

\begin{figure}[htbp]
	\centering
	\includegraphics[width=1\textwidth]{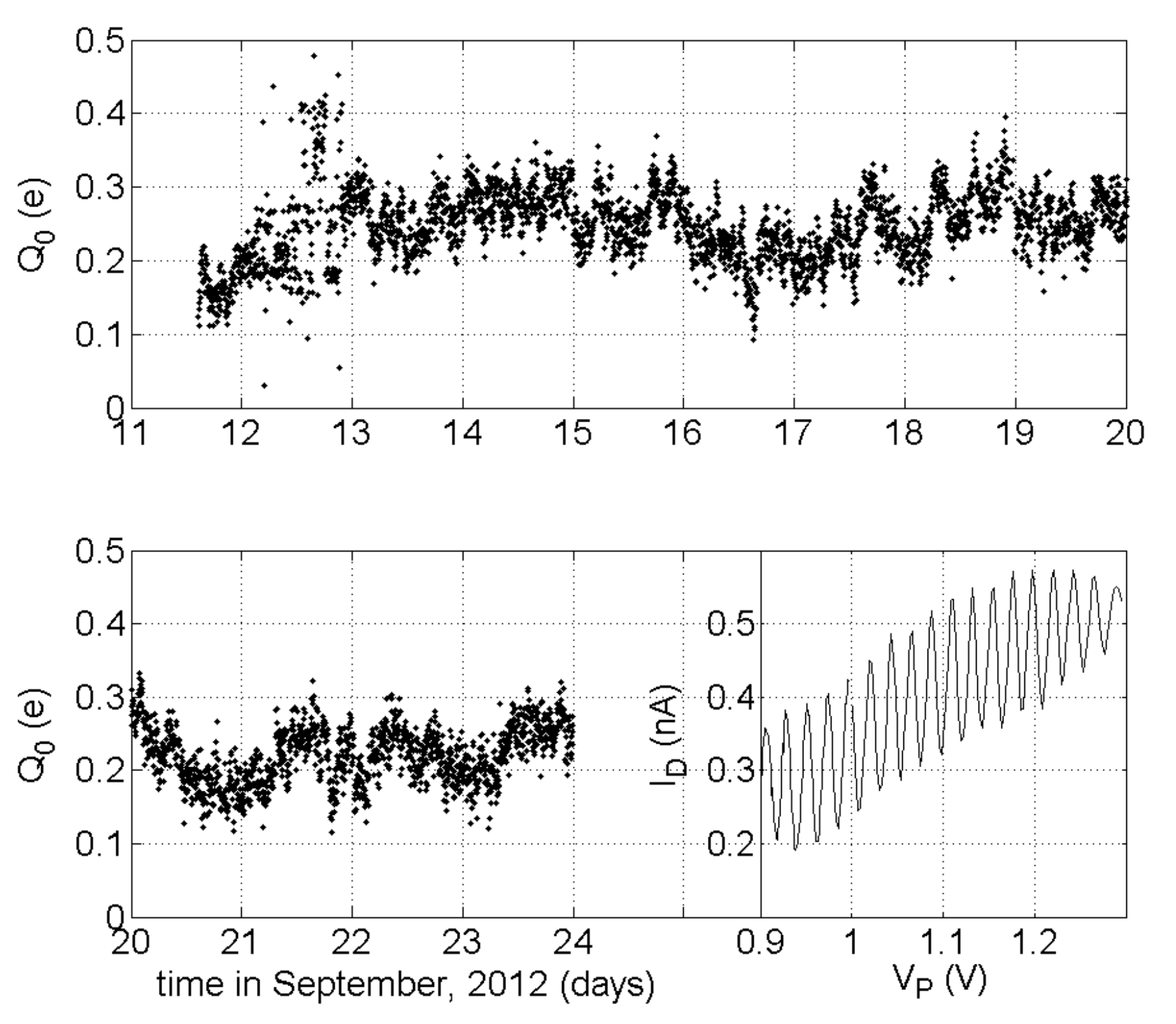}
	\caption[Figure 2]{
        Inset: Coulomb blockade oscillations in the SET.  The individual oscillations correspond
to adding one additional electron at a time to the quantum dot, and the overall monotonic
increase in the drain current reflects the smooth reduction of the tunnel barriers due to
the increase in \sub{V}{P}.  Main: charge offset drift \qot\ as a function of running
time, showing a range of about $\pm$ 0.15 $e$ overall, during the course of this 13 day
measurement.  Each individual data point was obtained by fitting a sinusoidal function
with a linear offset to the data as exemplified in the inset, for \sub{V}{P} between 1.1 and 1.2
V. $\sub{V}{D} = 0.5$ mV, $\sub{V}{L1} = \sub{V}{L2} = 1.4$ V, $\sub{V}{B1} = 0.296$ V,
$\sub{V}{B2} = 0.34$ V, $T = 2.2$ K.}
	\label{fig:I_VD_Q0}
\end{figure}


\begin{table}[htbp]
	\centering
	\includegraphics[width=1\textwidth]{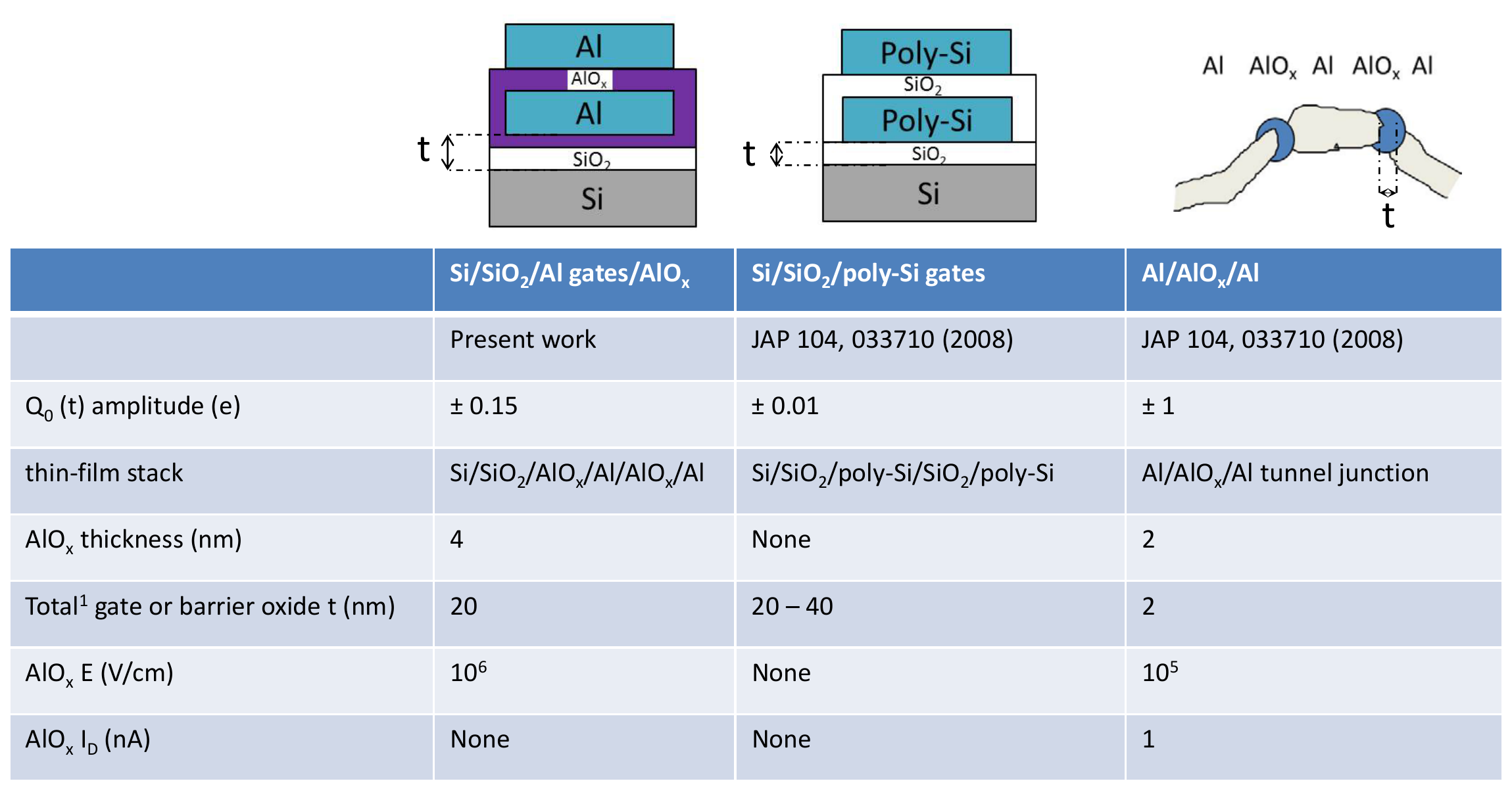}
	\caption[Table I]{
        Comparison of important attributes relevant to charge offset drift \qot\ for
three different classes of devices.  In the sketches, ``t'' represents the total gate or
barrier oxide thickness; leftmost sketch represents a region where there are two Al layers
above the wafer (e.g., where P and L1 overlap but not B1).  (i) For the Si/Al gate
devices, total t and electric field strength E refer to the SiO$_2$/AlO$_x$ between gate P
and dot.}
	\label{tab:comp}
\end{table}


%
%
\vfil\break
\bibliography{ref_SET,ref_QI_08}

%
%
%
%
%
%

\end{document}